\documentclass{appolb}
\usepackage{graphicx}
\usepackage{amsmath}
\usepackage{amssymb}
\usepackage[numbers, compress]{natbib}
\begin{document}
\title{Digital Radio Arrays for the Detection of Air Showers Initiated by Ultra-High-Energy Particles%
\thanks{Presented at XXVIII Cracow EPIPHANY Conference, January 2022}%
}
\author{Frank G.~Schr\"oder
\address{Bartol Research Institute, Department of Physics and Astronomy, University of Delaware, Newark DE, USA; and\\
Institute for Astroparticle Physics, Karlsruhe Institute of Technology (KIT), Karlsruhe, Germany}
}
\maketitle
\begin{abstract}
Digital radio arrays have become an effective tool to measure air showers at energies around and above 100 PeV.
Compared to optical techniques, the radio technique is not restricted to clear nights.
Thanks to recent progress on computational analysis techniques, radio arrays can provide an equally accurate measurement of the energy and the depth of the shower maximum. 
Stand-alone radio arrays offer an economic way towards huge apertures, e.g., for the search for ultra-high-energy
neutrinos, but still require technical demonstration on large scales. 
Hybrid arrays combining radio antennas and particle detectors have already started to contribute to cosmic-ray physics in the energy range of the presumed Galactic-to-extragalactic transition.
In particular, the combination of radio and muon detectors can pave a path to unprecedented accuracy for the mass composition of cosmic rays. 
This proceeding reviews recent developments regarding the radio technique and highlights selected running and planned antennas arrays, such as GCOS, GRAND, the SKA, the AugerPrime Upgrade of the Pierre Auger Observatory, and IceCube-Gen2.
\end{abstract}

\section{Introduction}
Radio detection of cosmic-ray air showers has matured as an additional technique for ultra-high-energy astroparticle physics. 
Following the success of several prototype experiments, a new generation of digital antenna arrays will be one of the drivers of the field in the coming decades. 
Depending on the science goals, these will be either stand-alone radio arrays or hybrid arrays combining radio and particle detectors, which enhances the measurement accuracy for air-shower parameters and, thus, for the properties of the primary cosmic particles. 

The success of the radio technique has its foundation, on the one hand, in technical advances, such as calibration techniques for timing \cite{PierreAuger:2015aqe,Schroder:2010sa} and amplitude \cite{LOPES:2015eya,PierreAuger:2017xgp,NellesLOFAR_calibration2015,Bezyazeekov:2015rpa}, and the development of dedicated analysis tools of digitally acquired radio data \cite{RadioOffline2011,Glaser:2019cws}.
On the other hand, an improved understanding of the radio emission mechanism has been critical to interpret and use the radio measurements for cosmic-ray science. 
Nowadays, state-of-the-art simulation codes \cite{HuegeCoREAS_ARENA2012,Alvarez_ZHAires_2012} reproduce and predict radio signals of air showers with sufficient accuracy to apply them for template matching \cite{BuitinkLOFAR_Xmax2014,Bezyazeekov:2018yjw} and machine learning techniques \cite{Erdmann:2019nie,Bezyazeekov:2020qqi,Rehman:2021nw} when analyzing radio data of air showers.

Last but not least, essential for the success of the radio technique is its intrinsic value for cosmic-ray measurements. 
Radio detectors are the ideal complement for arrays of particle detectors, in particular, muon detectors because the radio detectors are sensitive to the atmospheric depth of the shower maximum, $X_\mathrm{max}$, and the energy content of the electromagnetic component.
Those radio measurements combined with the muon content of the air shower provide maximum statistical sensitivity to the mass of the primary cosmic rays \cite{Holt:2019tja}.

\section{Characteristics of the Radio Emission of Air Showers}

Radio emission by air showers is coherent and beamed in the forward direction with a relatively small opening angle of at most a few degrees. 
At frequencies below a few $100\,$MHz, the radio emission is emitted in a cone that extents slightly beyond the Cherenkov angle, as the amplitude falls quickly at larger distances.
At higher frequencies, the coherence condition of the emission is fulfilled only at the Cherenkov angle and the footprint becomes a ring with a diameter given by the Cherenkov angle and the distance to the radio emission which is mostly from a region close to the shower maximum \cite{Glaser:2016qso}.

These general characteristics hold for two emission mechanisms: geomagnetic and Askaryan emission.
As published recently \cite{James:2022mea}, the nature of these emission mechanism depends on the frequency band and density of the medium.
Except for very inclined showers or at high frequencies of several $100\,$ MHz to GHz, the situation for air-showers is as follows:
the geomagnetic emission is mostly due to the induction of a transverse current by the Earth's magnetic field; it dominates over the Askaryan emission which is mostly due to the time variation of the negative net charge excess in the shower front.
The situation is completely different in dense media, such as ice, where the geomagnetic emission is negligible and the Askaryan emission can be understood as coherent Cherenkov-light emitted by a net charge that moves faster than the speed of light in that medium.

While the discussion of the emission principles is of pedagogic value, it is less important for the practical application, as modern simulation codes calculate the complete radio emission by all the electrons and positrons in an air shower.
They thus correctly predict superposition and polarization effects, e.g., an asymmetry for the radio lateral distribution due to the interference of geomagnetic and Askaryan emission.

Further important characteristic of the radio emission of air showers are: 
The radio footprint has a radius of $o(100\,$m) for vertical showers and few km around the shower axis for very inclined showers, which can extent to several $10\,$km on ground by projection effects along the shower direction \cite{AERAinclined_JCAP2018}.
At larger distances, the emission is strongest at several $10\,$MHz corresponding to wavelengths larger than the thickness of the shower disk.
At the Cherenkov angle it extents to several GHz \cite{NellesLOFAR_CherenkovRing2014,Smida:2014sia}. 
Generally, the frequency spectrum is broad band, corresponding to a short duration of a few ns to a few $100\,$ns of the radio pulse.
Due to the Galactic noise being the dominant radio background, the signal-to-noise-ratio is best above a few $100\,$MHz, at least when close to the Cherenkov ring \cite{BalagopalV:2017aan}.

Due to the combination of these properties, the optimum frequency band and antenna spacing of an array depends on the targeted range of zenith angles. 
For full sky coverage including vertical showers, a dense spacing of $o(100\,$m) is required and high detection frequencies above $100\,$MHz are best. 
Nonetheless, with larger antenna spacings of $1-2\,$km at lower frequency bands, such as $30-80\,$MHz, inclined air showers can be detected, and such arrays can provide a huge aperture for low cost.

\section{Reconstruction of the Properties of the Primary Cosmic Particles}
Most important for the practical application of the radio technique is the performance of antenna arrays for the reconstruction of air-shower parameters and, subsequently, the properties of the primary cosmic-ray particles. 
For stand-alone radio arrays, in addition, the detection efficiency needs to be well understood and recently some progress has been made \cite{Lenok:2021sW}.
Nonetheless, determining the radio efficiency is less important for hybrid arrays when the efficiency is determined by collocated particle detectors.
Therefore, I summarize here the status on the accuracy achieved for key shower parameters, as these are important for both cases, stand-alone and hybrid radio arrays.

\subsection{Arrival Direction}
\begin{figure}[t]
\centerline{%
\includegraphics[height=4.9cm]{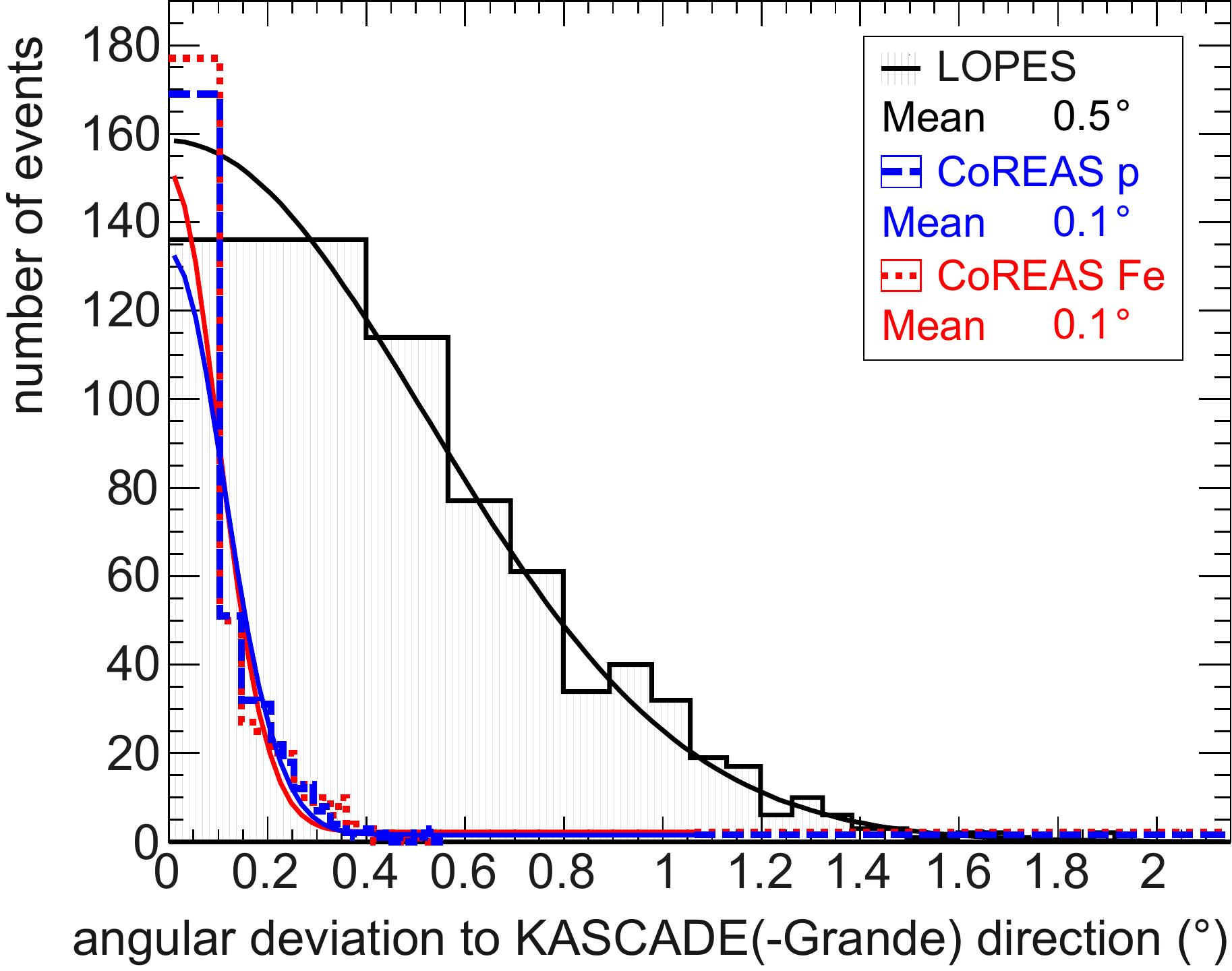}
\hfill
\includegraphics[height=4.9cm]{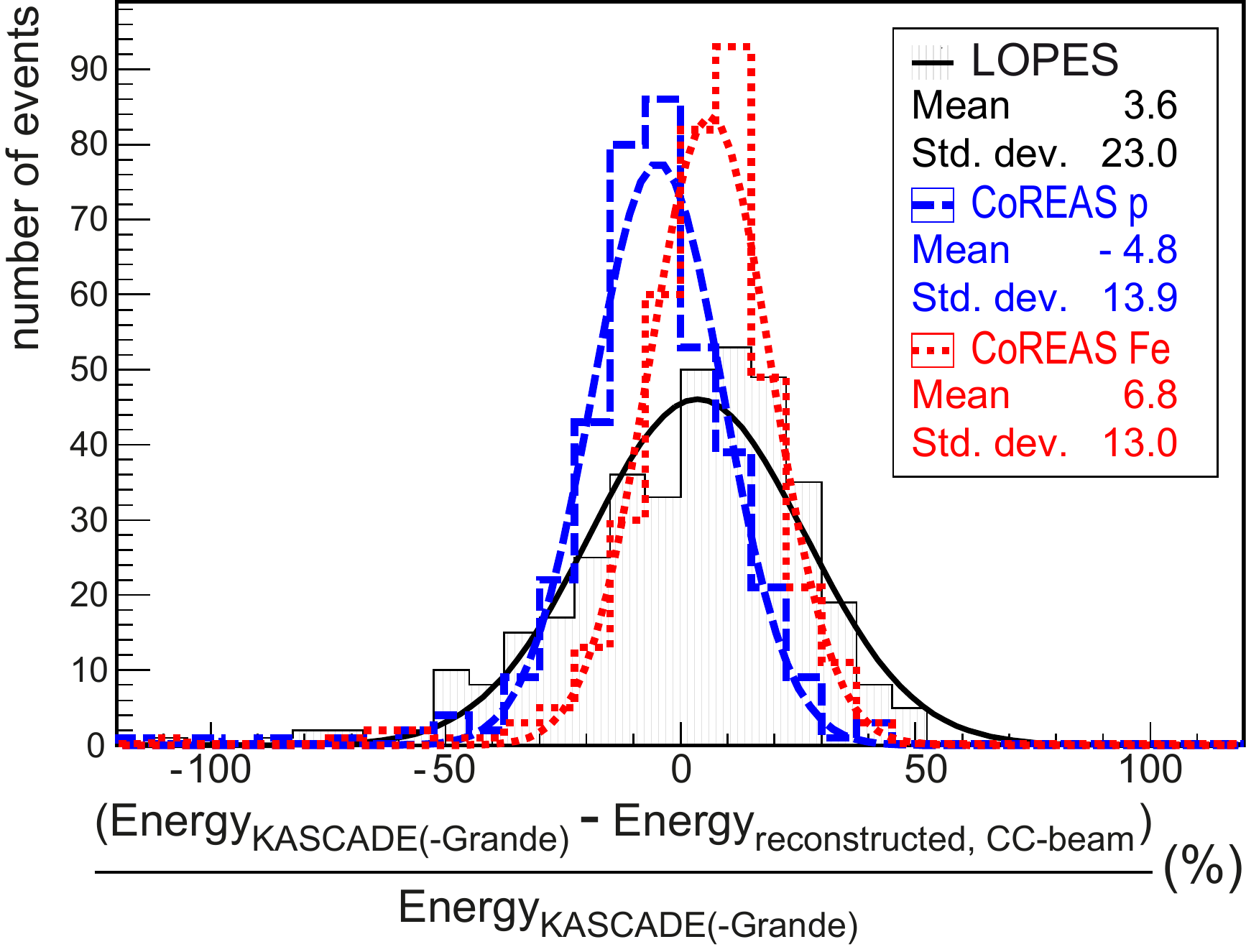}
}
\caption{Interferometric LOPES measurements as example for direction and energy reconstruction. The LOPES measurements are compared to the coincident measurements of the KASCADE and KASCADE-Grande particle detectors (black), where the spread contains the uncertainty of both measurements and constitutes an upper limit of the radio precision. 
Additionally, the deviation of CoREAS end-to-end simulations to the true input values of the simulations is shown to indicate the expected resolution of the radio measurements alone (Fig.~from Ref.~\cite{Apel:2021oco}).}
\label{Fig:LOPES}
\end{figure}

The arrival direction of air showers can be reconstructed with $o(1^\circ)$ accuracy  by triangulation through the pulse arrival times in at least three antennas of an array.
Subdegree resolution is possible when taking the hyperbolic shape of the radio wavefront \cite{Apel:2014usa,Corstanje:2014waa} into account that may be approximated by either a spherical or conical wavefront depending on the case.
Also digital interferometry provides a subdegree resolution of the arrival direction (Fig.~\ref{Fig:LOPES}) \cite{Apel:2021oco}.

\subsection{Energy}
Due to the coherent nature of the radio emission, its energy content scales quadratically with the size of the electromagnetic shower component, which itself is closely correlated with the energy of the primary particle. 
Therefore, the energy of the primary particle can be reconstructed in various ways based on measurements of the amplitude \cite{LOPES:2014bps,Tunka-Rex:2015zsa} or energy fluence \cite{AugerAERAenergy2015} either in individual antennas or by using the amplitude of a beam formed into the shower direction \cite{Apel:2021oco}.
Even a single antenna can be sufficient if the shower geometry is provided by a collocated air-shower array \cite{Tunka-Rex:2016gcn,Welling:2019scz}, however, with limited energy resolution.

At high signal-to-noise ratios, both, the precision and the absolute scale of the energy, are limited by systematic uncertainties, the largest currently being the calibration of the antennas. 
State-of-the-art antenna arrays achieve a total accuracy for the energy of the primary particle of slightly better than $15\,\%$, and $10\,\%$ seems in reach with improvements of the antenna calibration.
Radio antennas are thus also a promising tool to calibrate and compare the absolute energy scales of different cosmic-ray observatories \cite{Tunka-Rex:2016nto,Mulrey:2020oqe}.

\begin{figure}[t]
\centerline{%
\includegraphics[width=12.5cm]{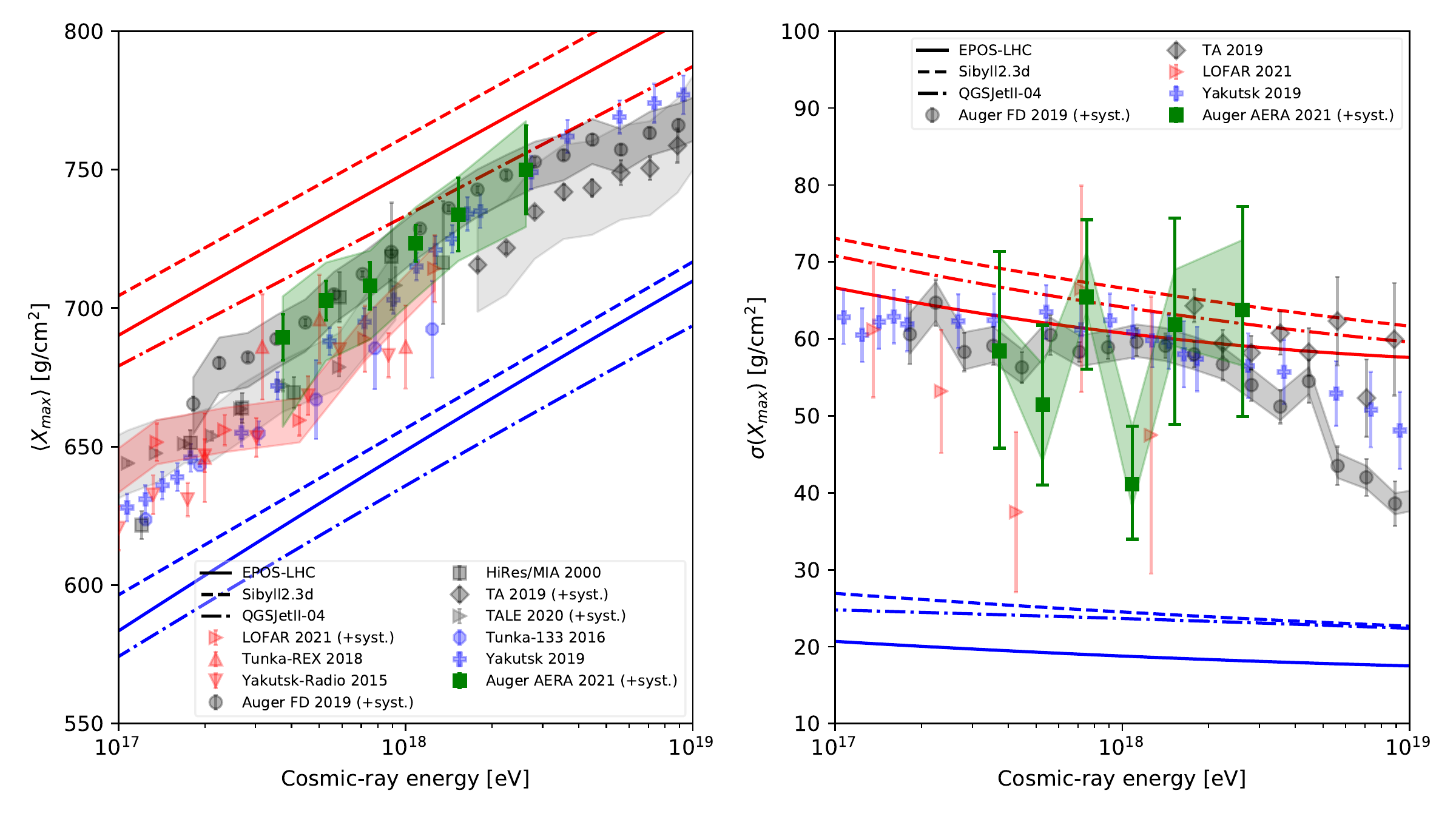}
}
\caption{$X_\mathrm{max}$ (left) and its standard deviation (right) measured by various experiments, among them a few radio arrays. The possible tension between LOFAR \cite{Corstanje:2021cC} and Auger \cite{Abreu:2021CB} is under investigation by a joint working group (courtesy of Bjarni Pont \cite{PontPhD2021}).}
\label{fig:XmaxByRadio}
\end{figure}

\subsection{Type and Mass of the Particle}
Radio arrays can contribute in a number of ways to the measurement of the type and mass of the primary particles. 
Photons, protons, and heavier nuclei initiate showers of different depth of the shower maximum and of different muon content. 
As these relations are of statistical nature, a precise measurement of the mass of an individual cosmic ray will be cumbersome if not impossible in most cases. 
Nonetheless, the mass composition can be assessed statistically and individual events can be classified into mass groups. 

A classical way is the measurement of the atmospheric depth of the shower maximum, $X_\mathrm{max}$ (Fig.~\ref{fig:XmaxByRadio}).
Several parameters of the radio signal are sensitive to $X_\mathrm{max}$, among them: the slope of the lateral distribution \cite{Tunka-Rex:2015zsa,LOPES:2012xou}, the shape of the wavefront \cite{Apel:2014usa}, the slope of the frequency spectrum \cite{Grebe:2013lvs}, the diameter of the Cherenkov ring, and the fraction of Askaryan relative to geomagnetic emission, which is accessible via the polarization of the radio signal \cite{Paudel:2022tbe}.
Moreover, $X_\mathrm{max}$ can be reconstructed by interferometric methods \cite{Apel:2021oco,Schoorlemmer:2020low,Schluter:2021egm}.
The highest accuracy of better than $20\,$g/cm$^2$ has been achieved by matching templates of CoREAS simulations against the measured radio pulses \cite{BuitinkLOFAR_Xmax2014,Bezyazeekov:2018yjw}. 

Nonetheless, even the mass resolution achievable by a perfect $X_\mathrm{max}$ measurement is limited intrinsically by shower-to-shower fluctuations.
Overcoming that limitation requires the simultaneous measurement of additional shower parameters, such as higher-order parameters of the longitudinal shower profile.
In particular, the width parameter $L$ is expected to be measurable with high precision at the ultra-dense Square Kilometre Array (SKA) planned in Australia \cite{Buitink:2021Mv}. 
This will open a new way to test hadronic and electromagnetic models of the shower development and eventually increase the mass resolution.

Another methods that can be applied in the near-term future is the combination of radio and muon measurements. 
Since the ratio between the radio amplitude and the muon density provides additional mass sensitivity only partly correlated with $X_\mathrm{max}$, coincident radio and muon measurements have the potential to achieve unprecedented accuracy for the measurement of the mass composition \cite{Holt:2019tja}.

\section{Selected Experiments}
This section introduces several recent and planned air-shower arrays making use of the radio technique, some of them as hybrid arrays together with particle detectors.
Focusing on the future of the digital radio technique, I skip over past efforts by simply providing some references for further reading. 
Pioneering digital radio arrays have been LOPES \cite{Apel:2021oco}, the radio extension of the KASCADE-Grande particle-detector array in Germany \cite{Apel:2010zz}, and CODALEMA in France \cite{Ardouin:2009zp}. 
They were followed by a second generation of radio arrays, such as AERA (see below), LOFAR \cite{SchellartLOFAR2013}, TREND \cite{Charrier:2018fle}, and Tunka-Rex \cite{Bezyazeekov:2015rpa}. 
Moreover, air-showers have been detected by balloon-borne radio detectors \cite{ANITA_CR_PRL_2010}.
Finally, various approaches have been tested and are developed to search for the radio emission of neutrino-induced showers in dense media \cite{ARA_PRD_2015,Barwick:2016mxm,RNO-G:2020rmc,BrayReview2016}.
For a summary on these experiments, I refer to a number of review articles \cite{Huege:2016veh,Schroder:2016hrv,Connolly:2016pqr,Sokolsky:2022yqj}.

\subsection{Radio Detection at the Pierre Auger Observatory}
The Pierre Auger Observatory is the largest air-shower array in the world. 
Its main detector systems are an array of water-Cherenkov detectors overlooked by fluorescence telescopes \cite{PierreAuger:2015eyc}.
In an enhancement area, the spacing of the surface detectors is denser and additional underground muon detectors are being installed \cite{PierreAuger:2020gxz}. 
That enhancement area also hosts the Auger Engineering Radio Array (AERA) comprised of more than $150$ antenna stations on $17\,$km$^2$ \cite{Huege:2019snr}.

Today, this is the largest radio array dedicated to cosmic-ray detection. 
Planned as an engineering array, several antenna types and data-acquisition systems have been tested, and a number of pioneering analyses have been performed, e.g., on energy \cite{PierreAuger:2016vya} and $X_\mathrm{max}$ reconstruction \cite{Abreu:2021CB} including a thorough study of systematic uncertainties.
As a result, the accuracy can compete with the fluorescence telescopes. 
Nonetheless, radio antennas cannot fully replace the fluorescence telescopes because with a similarly sparse spacing as the water-Cherenkov detectors, radio arrays can observe only very inclined showers with $\theta \lesssim 60^\circ$ with full efficiency. 

\begin{figure}[t]
\centerline{%
\includegraphics[height=6cm]{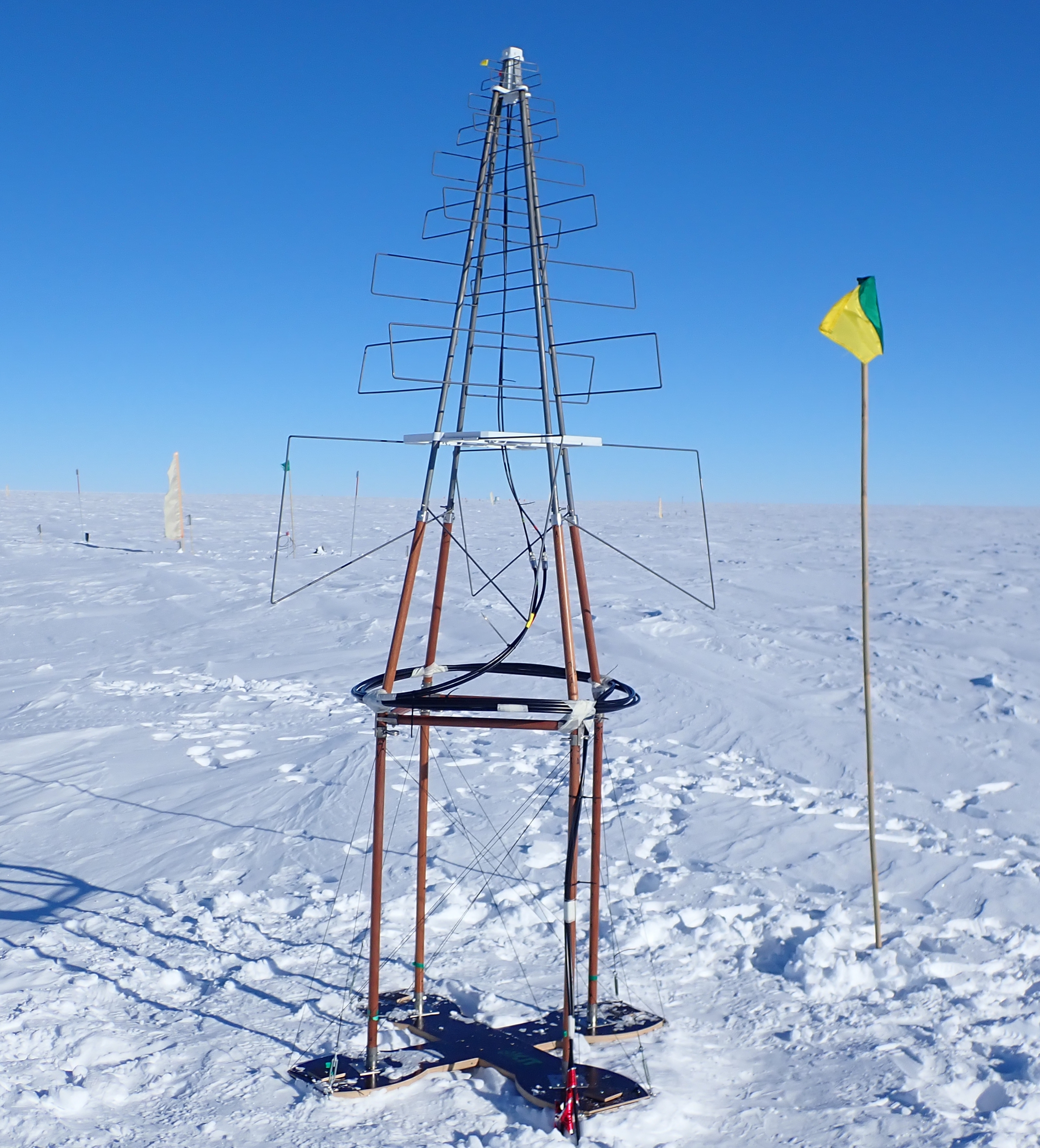}
\hfill
\includegraphics[height=6cm]{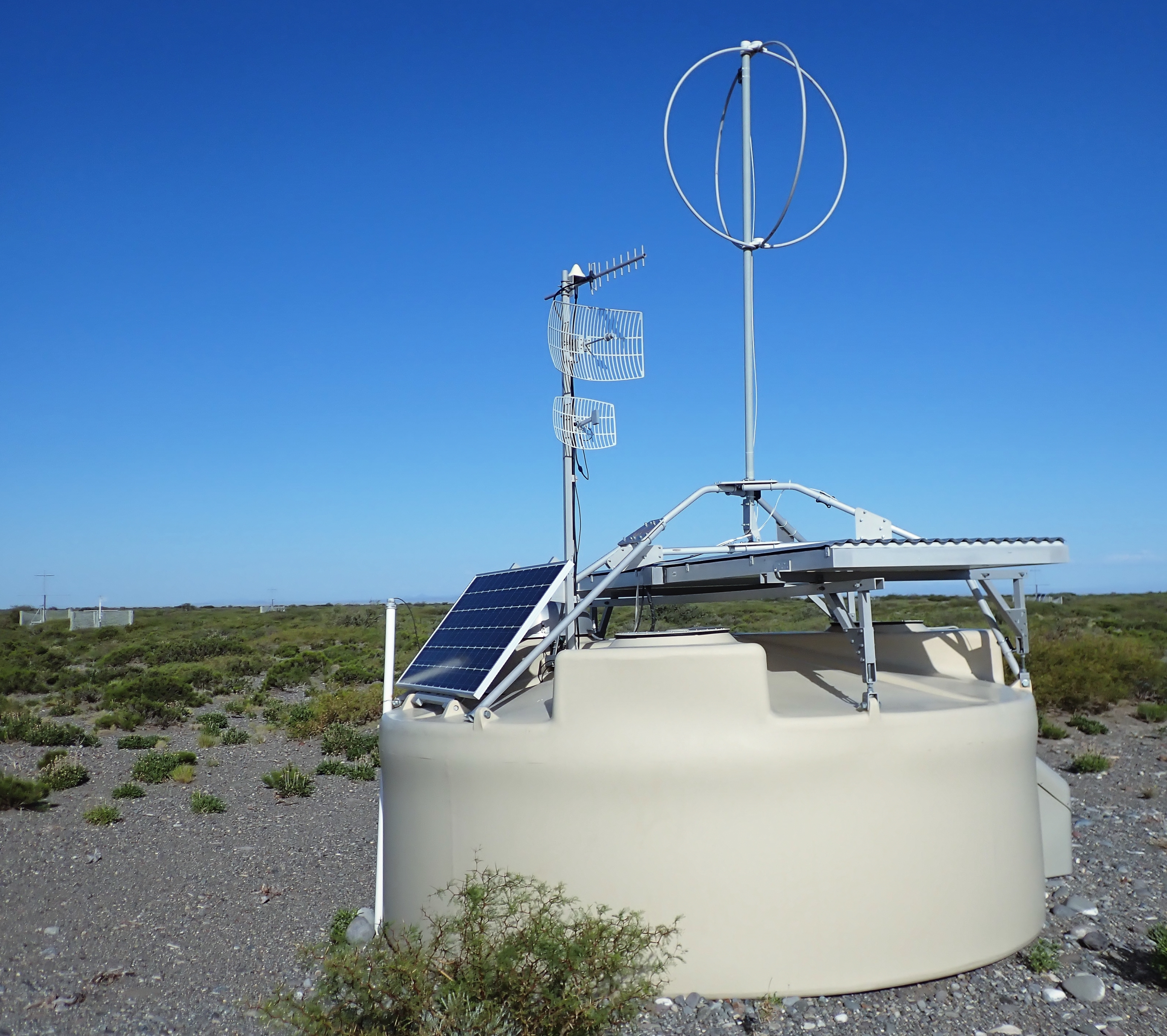}
}
\caption{Left: SKALA-type radio antenna installed at the South Pole as part of the prototype station of IceCube's surface enhancement. Right: AugerPrime surface detector consisting of a water-Cherenkov detector upgraded by scintillation panel and a SALLA-type radio antenna.}
\label{Fig:Antennas}
\end{figure}

Targeting these inclined air showers, radio antennas are added to the ongoing AugerPrime upgrade \cite{RadioAuger_ICRC2021}. 
The water-Cherenkov detectors will be enhanced not only by a scintillation panel on top, but also by a radio antenna of SALLA type (Fig.~\ref{Fig:Antennas}~right) -- a more advanced version of the SALLA that was developed as a candidate antennas for AERA and afterwards used at Tunka-Rex \cite{PierreAuger:2012ker,Bezyazeekov:2015rpa}.
For the very inclined showers, the radio antennas will provide a cost-effective measurement of the electromagnetic component. 
Combined with the muonic component measured by the water-Cherenkov detectors, this allows for further investigation of the muon puzzle and for mass discrimination.

\subsection{IceCube's Surface Enhancement and IceCube-Gen2}
By its combination of the IceTop surface array of ice-Cherenkov detectors with a deep optical detector in the Antarctic ice, IceCube is a unique lab for cosmic-ray air showers. 
To mitigate the effect of snow accumulation on IceTop, an enhancement array of elevated scintillation panels and radio antennas is planned. 
The radio antennas will increase the accuracy for the electromagnetic component of the air showers and, thus, for the overall accuracy for the identification of photons and separation of mass groups of cosmic rays \cite{Schroder:2019suq,Haungs:2019ylq}.

A complete prototype station of eight scintillation panels and three SKALA-type radio antennas (Fig.~\ref{Fig:Antennas}~left) \cite{SKALAv2} was installed at the South Pole in 2020, and has successfully detected air showers in coincidence with IceTop \cite{IceCube:2021epf}.
Further prototype stations will be installed at the Pierre Auger Observatory in Argentina and the Telescope Array in Utah, which will enable additional cross-checks with their particle-detector arrays. 
In particular, by the direct comparison it can be directly tested whether and to what extent the high frequency band of the SKALA antennas (up to about $350\,$MHz) can lower the detection threshold compared to $30-80\,$MHz band currently used at the Pierre Auger Observatory.
At the same time, preparation are continuing for the deployment of a full enhancement array at IceTop.

\begin{figure}[t]
    \centering
    \includegraphics[width=0.59\linewidth]{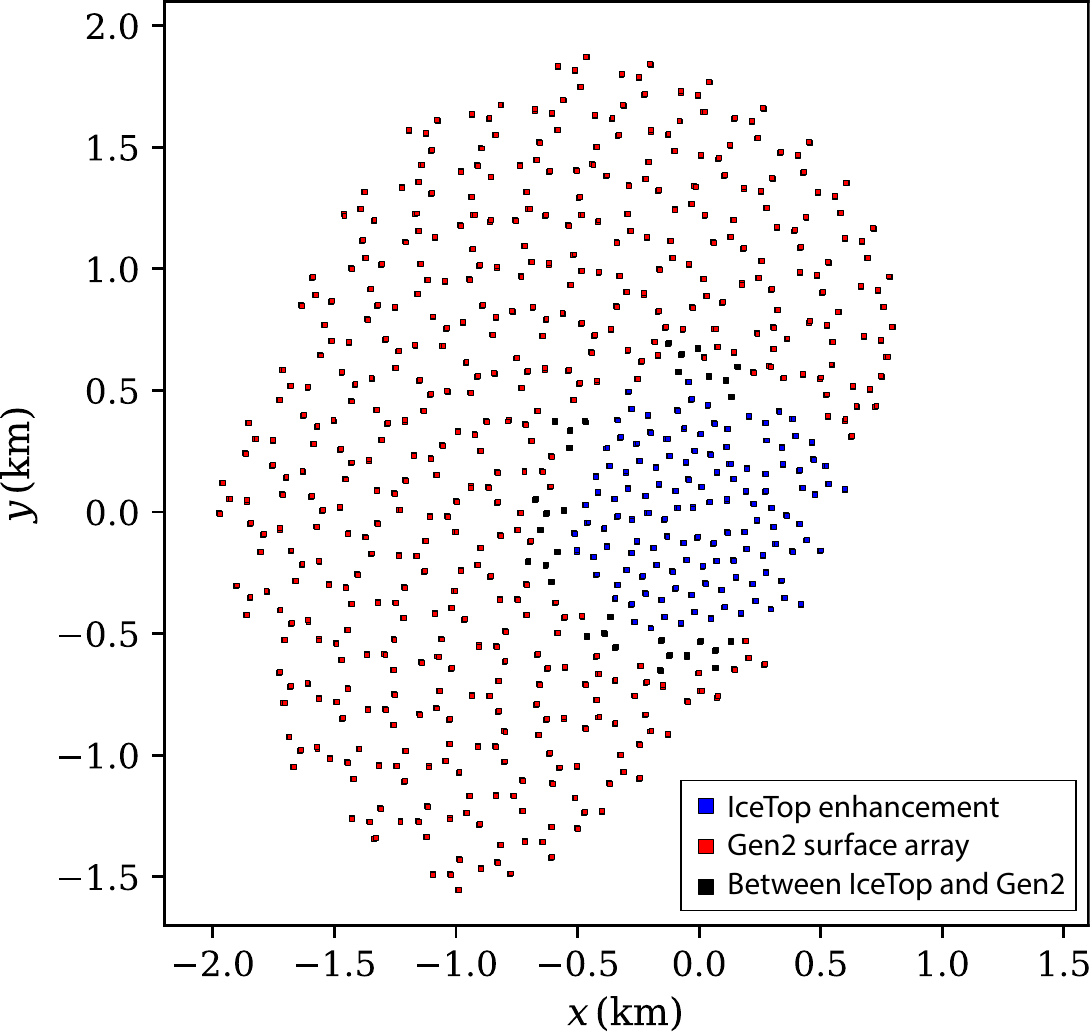}
    \hfill
    \includegraphics[width=0.39\linewidth]{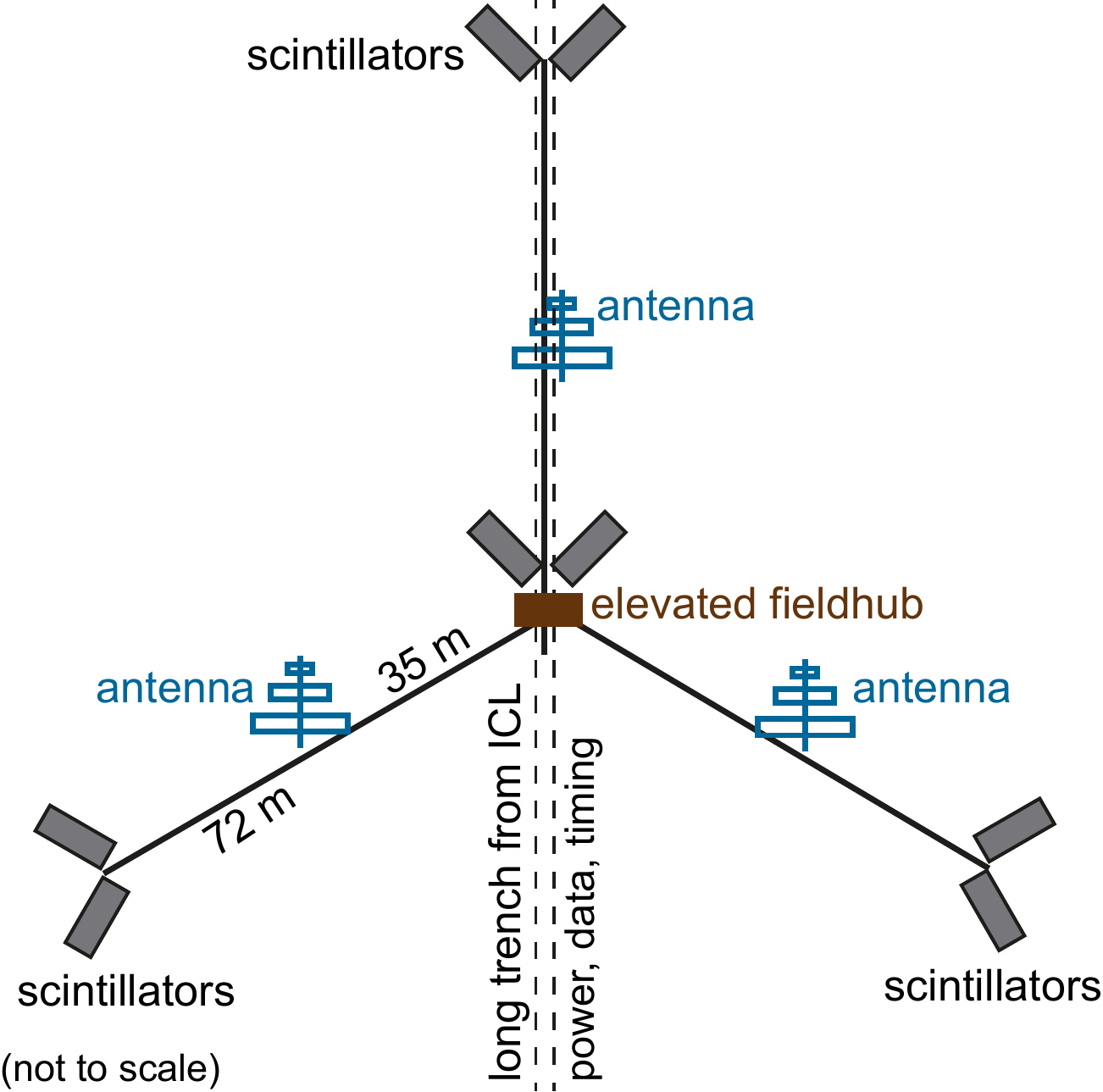}
    \caption{Planned layout of the surface array of IceCube-Gen2 (left) and the design of an individual surface station design (right), which is comprised of eight scintillation panels and three elevated radio antennas. The different colors indicate stations over the existing deep IceCube detector and its planned IceCube-Gen2 extension by additional 120 optical strings (Fig.~from Ref.~\cite{IceCubeGen2surface_ICRC2021}).
    }
    \label{fig:layout}
\end{figure}

As the design has demonstrated to detect air showers, it has also been selected as baseline design for the IceCube-Gen2 surface array \cite{IceCube-Gen2:2020qha,IceCubeGen2surface_ICRC2021}.
This will increase the total size of the combined scintillator and radio array to about $6\,$km$^2$ on top of the enlarged optical in-ice array of IceCube-Gen2.
The benefit is not only the simple increase in surface area, but additionally the sky range for those events with coincident detection in the surface and deep detectors will significantly increase.
This results in a more than 30-fold higher aperture over IceCube for these unique air-shower events, which enables a number of science cases, such as the study of prompt decays in air showers or high precision for the separation of mass groups.
Consequently, IceCube-Gen2 will be a leading detector to study air-shower particle physics and cosmic-ray physics in the energy range up to a few EeV.

\subsection{Giant Radio Array for Neutrino Detection (GRAND)}
The Giant Radio Array for Neutrino Detection (GRAND) is planned as a huge multi-site project covering $200,000\,$km$^2$ with a density of one antenna station per square kilometer \cite{GRAND:2018iaj}.
Although optimized for the detection of air showers initiated by tau leptons produced in neutrino interaction, GRAND will also be an excellent detector for very inclined cosmic rays.
As a first step, GRANDProto300 with $300$ antennas will be deployed as a pathfinder in China followed by GRAND10k of $o(10,000)$ antennas. 

A particular challenge will be a fully efficient self-trigger for the autonomous stations of GRAND. 
While self-triggering itself has been demonstrated \cite{RAugerSelfTrigger2012,Charrier:2018fle}, improvements are needed with respect to efficiency and purity.
Featuring also a collocated particle-detector array, GRANDproto300 will be able to provide that demonstration and also contribute to cosmic-ray science in the energy range of the Galactic-to-extragalactic transition.

\subsection{Global Cosmic Ray Observatory (GCOS)}
The Global Cosmic Ray Observatory (GCOS) is another huge air-shower array planed to drive the field in the coming decades \cite{GCOS_ICRC2021}. 
GCOS may be smaller than GRAND, but as hybrid array of particle detectors and possible radio antennas as well as fluorescence telescopes, it aims for maximum accuracy on the measured cosmic-ray air showers. 
In the same way as for AugerPrime, the combination of radio antennas and water-Cherenkov tanks can provide mass sensitivity for inclined air showers. 
Thus, GCOS with its higher accuracy for the primary mass is an ideal complement to GRAND 
which aims at maximum exposure. 
Moreover, GCOS and GRAND could share one or several sites to save cost on infrastructure and to allow for cross-calibration between both arrays.

\section{Conclusion}
Summarizing, digital radio arrays are expected to play a major role in ultra-high-energy cosmic-ray physics during the next decades.
IceCube-Gen2 with the radio antennas of its surface array and the SKA will provide new insights into air-shower physics in the energy range up to a few EeV and target the astrophysical question of the transition from Galactic to extragalactic sources in this energy range \cite{SchroderAstro2020}.
For the highest energies, AugerPrime will pioneer hybrid detection of radio antennas and water-Cherenkov detectors measuring the electromagnetic and muonic components of very inclined showers. 
In the coming decades, GRAND with its huge radio arrays and GCOS comprised of hybrid arrays will be the leading ultra-high-energy air-shower arrays, searching for new fundamental physics at EeV to ZeV energies and for the sources of the most energetic particles known in the universe \cite{SarazinAstro2020}.


\section*{Acknowledgement}
This project has received funding from the European Research Council (ERC)
under the European Union’s Horizon 2020 research and innovation programme
(grant agreement No 802729).

\bibliography{references}

\end{document}